\documentclass[a4paper,12pt]{article}

\usepackage{amssymb}
\usepackage{graphicx}
\usepackage{url}
\urldef{\mailsa}\path|{yaflalo,nastyad,ron}@cs.technion.ac.il|
\usepackage{float}
\usepackage{subfigure}
\usepackage{listings}
\usepackage{graphicx}
\usepackage{epstopdf}
\usepackage{graphics}
\DeclareGraphicsExtensions{.jpg,.eps,.pdf,.png}
\usepackage{amsmath}
\usepackage{amsfonts}
\usepackage{enumerate}
\usepackage{textcomp}
\usepackage{color}
\usepackage{algorithmic}
\usepackage{algorithm}
\usepackage{bm}
\usepackage{dsfont}
\usepackage{color}
\newtheorem{defi}{Definition}[section]

\newcommand{\trace}{\operatorname{trace}}

\newcommand{\F}{\mathbf{F}}

\newcommand{\W}{\mathbf{W}}
\newcommand{\D}{\mathbf{D}}
\newcommand{\K}{\mathbf{K}}
\newcommand{\X}{\mathbf{X}}

\newcommand{\p}{\mathbf{P}}

\newcommand{\I}{\mathbf{I}}

\newcommand{\PHI}{\mathbf{\Phi}}

\newcommand{\A}{\mathbf{A}}

\newcommand{\LAMBDA}{\bm{\Lambda}}

\newcommand{\Alpha}{\bm{\alpha}}

\newcommand{\aalpha}{\bm{\alpha}}
\newcommand{\tr}[1]{\ensuremath{\text{ trace}\left(#1\right)}}

\newcommand{\ones}{\mathds{1}}

\newenvironment{disarray}%
 {\everymath{\displaystyle\everymath{}}\array}%
 {\endarray}
\everymath{\displaystyle\everymath{}}

\usepackage{xspace}
\usepackage{dsfont}
\makeatletter
\DeclareRobustCommand\onedot{\futurelet\@let@token\@onedot}
\def\@onedot{\ifx\@let@token.\else.\null\fi\xspace}

\def\etal{\emph{et al}\onedot}
\makeatother

\begin{document}


\title{Spectral Generalized Multi-Dimensional Scaling} 

\author{Yonathan Aflalo and  Anastasia Dubrovina {\upshape and} Ron Kimmel \\ Technion University, Haifa 3200, Israel}




\maketitle

\begin{abstract} 
Multidimensional scaling (MDS) is a family of methods that embed a
 given set of points into a simple, usually flat, domain.
The points are assumed to be  sampled from some metric space,
 and the mapping attempts to preserve the distances between each 
 pair of points in the set.
Distances in the target space can be computed analytically in this setting.
Generalized MDS is an extension that allows mapping one metric space into
 another, that is, multidimensional scaling into target spaces in which
 distances  are evaluated numerically rather than analytically.
Here, we propose an efficient approach for computing such mappings
 between surfaces based on their natural spectral decomposition,
 where the surfaces are treated as sampled metric-spaces.
The resulting {\em spectral-GMDS} procedure enables efficient embedding
 by implicitly incorporating smoothness of the mapping into the problem, 
 thereby substantially reducing the complexity involved in its solution 
 while practically overcoming its non-convex nature.
The method is compared to existing techniques that compute dense
 correspondence between shapes.
Numerical experiments of the proposed method demonstrate its efficiency 
 and accuracy compared to state-of-the-art approaches.
\end{abstract}

\section{Introduction}
Matching non-rigid or deformable shapes is a challenging problem
 involving a large number of degrees of freedom.
While matching rigid objects one needs to search for isometries in a three 
 dimensional Euclidean space, a problem that can be described
  by six parameters.
Matching solvers for rigid surfaces in $\mathbb{R}^3$ are known as iterative closest 
 point algorithms or ICP  \cite{chen1991object,besl1992method}. 
Non-rigid matching usually involves much more dimensions that can add up to the 
 number of points of the sampled surfaces that one wishes to match.
When ignoring the continuity and thus smoothness of matching one surface to another, 
 the problem can  be viewed as a combinatorial one, for which the computational
 complexity is exponential.
The  problem in this setting is NP hard,  which is the hardest to
 solve in terms of computational complexity.
The question we address is how to efficiently solve this notoriously
 hard problem.

Various attempts to define robust and invariant meaningful measures by which articulated 
 objects and deformable shapes could be identified were made. 
Adopting tools from metric geometry, the Gromov-Hausdorff distance 
 \cite{Gromov81,burago2001course},  
 and its variants were suggested as candidates for measuring the discrepancy 
 between two deformable shapes \cite{memoli,GMDS,bronstein2010gromov,raviv:bro:bro:kim:IJCV09}.
The Gromov-Hausdorff distance between two surfaces $S$ and $Q$, or $d_{\text{GH}}(S,Q) $ 
 in short, is the maximal distortion introduced when bijectively embedding $S$ into $Q$ and vice-versa. 
Motivated by early attempts of finding a common 
 parametrization for surfaces \cite{Schwartz:1989,Zigelman_2002},
 the idea of treating surfaces as metric spaces that can be embedded into simple spaces
 was  first suggested in \cite{Elad03}.
There, the metric of each surface is first embedded
  into a small dimensional Euclidean space, say $\mathbb{R}^3$, by a procedure
  known as multidimensional scaling \cite{MDS_review_book:1997,Rosman:2008,multi_grid_mds_2006}.
The flat mappings or {\it canonical forms} in the Euclidean space 
 are then treated as rigid surfaces and matched, for example, by ICP. 
Though the idea is appealing as far as its simplicity has to do, 
 the embedding error of mapping a non-flat manifold into a flat finite dimensional domain 
 can be substantial with little hope for convergence.
The question the geometric processing community was occupied  with, is how to avoid 
 intermediate simple spaces while still being able to computationally handle the seemingly 
 complicated task of matching non-rigid surface.
Towards that end,
 Memoli and Sapiro \cite{Memoli05,memoli} provided the support that sampling  surfaces 
 could be tolerated within the Gromov-Hausdorff framework.
In other words, the sampling error  is linear 
  as a function of the distance between the sampled points, 
  and could thus be bounded when comparing sampled surfaces. 
Equipped with that encouraging result,
 Bronstein \etal \cite{bronstein2006efficient} exploited the fact that the  $d_{\text{GH}}$ could be formalized 
 as three coupled generalized multidimensional scaling problems for which 
 they introduced a numerical solver \cite{GMDS}.

In retrospective,
 the Hausdorff measure optimized for by 
 celebrated iterative closet point  (ICP) procedure
  \cite{chen1991object,besl1992method,mitra2004registration}
  can be interpreted as a Gromov-Hausdorff distance 
  where distances are computed in the embedding $\mathbb{R}^3$ Euclidean space.
Other simple intermediate embedding spaces for matching non-rigid shapes  were advocated. 
The eigenspace of the Laplace-Baltrami operator was suggested in various flavors,
 for example by 
 Mateus \etal \cite{mateus2008articulated}, and by Rustamov \cite{Rustamov13}, 
 as potential Euclidean target space, 
 see also \cite{Berard1994,Coifman_Lafon:2006,levy2006laplace}. 
Lipman et al. \cite{mobius,lipman} embedded shapes conformally into disks 
 between which the correspondence boils down again to a six parameters M\"{o}bius transform,
 see also \cite{gu2004genus,jin2004optimal,zeng2012computing}. 
In that case, metric embedding errors are replaced by numerical ones, as important features 
 with effective Gaussian curvature often scale down substantially
 and can practically vanish when sub-sampled.
Partial remedy to this conformal distortion was proposed in \cite{aflalo2013conformal}. 
Kim \etal \cite{blended_map}  suggested a refinement procedure, while
 using conformal mappings that perform well only locally.
They softly tailored 
 a handful of such locally good maps, using a procedure they coined as {\it blending}. 

Heuristics that reduce the complexity of  
 the dense matching problem and detect some initial state at a significant 
  basin of attraction for convex solvers to refine were often employed  
  by the above approaches.
Such heuristics use feature point detectors and descriptors.
Some examples include the heat kernel signature (HKS) \cite{HKS,gebal2009shape}, 
  global point signature (GPS) \cite{Rustamov13}, 
  wave kernel signature (WKS) \cite{aubry2011wave}, 
  and scale-space representation \cite{zaharescu2009surface}.
Matching the metric spaces 
 with either geodesic \cite{Memoli05,GMDS} 
 or diffusion \cite{Berard1994,Coifman_Lafon:2006,bronstein2010gromov} distances, 
 could then be treated as a regularization or refinement term.
It produced dense correspondence
  from the sparse one provided by matching the feature points \cite{nastiaK11}.
Higher order structures were suggested for example in \cite{zeng2010dense}.
Dense matching was further accelerated by hierarchical solvers
 like \cite{sahillioglu2011coarse,raviv2012hierarchical}.
Still, the complexity of searching over the space of all possible point-to-point correspondences 
 was determined by the number points one wishes to match.

Ovsjanikov \etal \cite{fmap} illuminated the fact that given two functional spaces,
 and given the correspondence between these two spaces, there is a linear relation
 between the functional representation of a function in one space (shape) and its
 corresponding  functional representation in the second space (shape).
This linear relation is due to the given correspondence
 and can be viewed as a matrix translating the decomposition coefficients 
 of a function in one metric space to its set of corresponding coefficients in the other.
When the functional spaces are the eigenfunctions of the surface 
 LBO \cite{levy2006laplace}, the right matching matrix for isometric surfaces 
 would be nothing but the identity.
Ovsjanikov \etal \cite{fmap} named these linear connections between functional spaces
 as \emph{functional maps}  and used them to
 find dense correspondence between shapes. 
Under the assumption of \emph{smooth} function representation, for which
 the Laplace-Beltrami provides a natural basis   \cite{regularized_PCA}, 
 only a small number of leading eigenfunctions may be considered.
Thus, the combinatorial problem of correspondence detection can be casted
 as low dimensional functional map identification.
As always, in \cite{fmap,pokrass}, a number of matching regions or feature points 
 was required for computing the correspondence using functional maps.

When matching non-isometric shapes, the corresponding Laplace-Beltrami eigenspaces
 are incompatible. 
This effect is substantial, for instance, in the case of various human body 
 shapes in the SCAPE dataset \cite{anguelov2004correlated}.
To overcome that limitation, Kovnatsky \etal \cite{quasi_harmo} suggested constructing
 common approximate harmonic bases for pairs of shapes by joint diagonalization.
That is, the functional map, treated as a matrix, is restricted to be diagonal.
Pokrass et al. \cite{pokrass} subsequently formulated the non-rigid isometric 
 matching problem as permuted sparse coding.
There, the dense correspondence  
 is extracted through coupling the functional map representation 
 with that of matching corresponding regions.
The computation is performed by alternating minimization over the unknown functional 
 map, while penalizing non-diagonal solutions, and a permutation matrix, 
 representing the correspondences.

In this paper, we argue that the  $L_2$ version of the 
 Gromov-Hausdorff framework for matching 
 deformable shapes can be naturally casted into the spectral domain with a novel 
 functional map representation.
Here, we overcome the compromise of having to 
 match multiple semi-local or differential structures also known as sparse  matching, 
 while at the same token, reduce the overall complexity of the dense matching problem.
We utilize the following important observations:
\begin{itemize}
  \item The point-to-point correspondence itself between two shapes 
         can be thought of as a functional map between the functional spaces of the shapes.
  \item Distances measured on a shapes are smooth functions, and as such
   are well suited for our functional map representation.
    See recent theoretical and empirical support for the multidimensional scaling 
    case in  \cite{spectral_MDS}.
\end{itemize}  
We present a spectral formulation for the {\it generalized multidimensional scaling} method
  \cite{GMDS}, that we denote as {\it spectral GMDS}, or S-GMDS in short.
We show that the suggested procedure outperforms state-of-the-art 
 dense correspondence solvers in terms of complexity and accuracy 
  while substantially reducing the amount of required supporting features. 

\section{Notations}
We consider a two dimensional parametrized Riemannian manifold $M$,
 equipped with a metric tensor $G$.
The metric $G$ induces several scalar products $\langle\cdot,\cdot\rangle_G$.
\begin{itemize}
\item For any tangent plane of $M$ at any point $p\in M$, denoted by $T_p(M)$,
 given two vectors $(u,v)\in T_p(M)$,  $\langle u,v\rangle_G$ is defined by
$$
\langle u,v\rangle_G =  u^TGv.
$$
\item For any two  functions, $f$ and $h$, defined on $M$, $\langle f,h\rangle_G$ is defined as
$$
\langle f ,h \rangle_G = \iint_{p(M)} f(x)h(x)\sqrt{g}dx,
$$
where $p(M)$ represents the parametrization space of $M$,  and $g=\det(G)$.
\item For any two vector fields, $U$ and $V$, on $T(M)$,
 $\langle U,V\rangle_G$ is defined as
$$
\langle U,V\rangle_G=\iint_{p(M)} U(x)^T G V(x) \sqrt{g}dx.
$$

\end{itemize}
All the above scalar products induce their respective norms $\|\cdot\|_G=\sqrt{\langle\cdot,\cdot\rangle_G}$.
Finally, the metric tensor $G$ induces two differential geometric operators for any function $f$ defined over $p(M)$,
\begin{itemize}
\item $\nabla_G f=G^{-1} \nabla_x f = \sqrt{g}\sum_jg^{ij}\partial_j f,$
   where $g^{ij}=\left(G^{-1}\right)_{i,j}$ and $\partial_i$ is the derivative with respect to the $x_i$ coordinate.
\item
  $ \Delta_G f=\frac{1}{\sqrt{g}}\sum_i\partial_i\left ( \nabla_G f\right )
                     =\frac{1}{\sqrt{g}}\sum_i\partial_i\left(\sqrt{g}\sum_jg^{ij}\partial_j f\right).$
\end{itemize}

\section{Functional maps}
\label{sec:functional_maps}
Given two shapes $S_1$ and $S_2$, a functional map between
 $S_1$ and $S_2$ maps any function $f_1:S_1\rightarrow \mathbb{R}$
 to its image $f_2:S_2\rightarrow \mathbb{R}$.
This map could be represented by an operator $\mathcal{K}$ defined on the functional
 space $\left\{f_1:S_1\rightarrow \mathbb{R}\right\}$ and obtaining its
 values in $\left\{f_2:S_2\rightarrow \mathbb{R}\right\}$,
 such that $f_2=\mathcal{K}(f_1)$.
If the mapping is linear, $\mathcal{K}$ is a linear operator and can be defined through
 a kernel $k:S_1 \times S_2\rightarrow \mathbb{R}$, where
\begin{equation}
\label{eq:continuous_func_map}
  f_2(y)=\mathcal{K}\left[f_1\right](y) = \int_{S_1}k(x,y)f_1(x)da_1(x),
\end{equation}
 where $x \in S_1,\,\, y \in S_2$, and $da_1(x)=\sqrt{g_1}dy$, here $g_1 = \det{(G_1)}$,
  represents an infinitesimal area element of $S_1$.
For every kernel $\mathcal{K}$, we define its conjugate $\mathcal{K}^*$ as
\begin{equation}
 \label{eq:conjugate_map}
\mathcal{K}^*\left[f_2\right](x) = \int_{S_2}k(x,y)f_2(y)da_2(y).
\end{equation}
For simplicity, consider $S_1$ and $S_2$ to be two triangulated surfaces,
 in which case, a discrete version of (\ref{eq:continuous_func_map}) can be defined
 by a matrix $\K$, such that
\begin{equation}
 \label{eq:discrete_func_map}
 f_2=\K\A_1f_1.
\end{equation}
Here, $\A_1$ is a diagonal matrix in which $\{A_1\}_{ii}$ is the area of the Voronoi cells about
 vertex $i$ as introduced in \cite{cotan},
 and $\K_{i,j}=k(x_i,y_j)$.

\subsection{Properties of functional maps}
\label{subsec:Charact_func_map}
A functional map, linearly relating two functional spaces, represents
 an arbitrary relation between the two spaces.
In order for such a mapping to have a practical meaning
  we  need some constraints that would restrict it  to a subspace of
  the possible functional maps from $S_1$ to $S_2$.
Specifically, we require the following properties
\begin{enumerate}
\item Linearity, $\mathcal{K}(f+\lambda h)=\mathcal{K}(f)+\lambda\mathcal{K}(h)$.
\item Smoothness, $\mathcal{K}$ should map a smooth function to a smooth function.
\item Unitarity, $f=\mathcal{K}^*\left(\mathcal{K}(f)\right)$.
\item Mass preservation,
\begin{eqnarray*}
 \int_{S_2}\mathcal{K}(f_1)da_2 & =  \int_{S_1}f_1da_1
 \\
 \mbox {\,\,\, and  \,\,\,} 
 \\
  \int_{S_1}\mathcal{K}^*(f_2)da_1& =\int_{S_2}f_2da_2.
  \end{eqnarray*}
  \item Local area preservation,  or  generalized Parseval's identity,
  \begin{eqnarray*}
  \forall \Omega \subseteq S_1, \int_{S_1} \ones_\Omega da_1=\int_{S_2} \mathcal{K}(\ones_\Omega) da_2.
  \end{eqnarray*}
 where $\ones_\Omega$  is an indicator function that is equal to one in  $\Omega$ and zero elsewhere.
  \item Conformality, if $(u,v)$ is a conformal parametrization of $S_1$, then
   $( \mathcal{K}(u), \mathcal{K}(v))$ is a conformal parametrization of $S_2$.
\end{enumerate}
  It is shown in \cite{Rustamov13} that,
  \begin{itemize}
  \item Local area preservation   holds, if and only if,\\ 
  $\int_{S_1} hfda_1=\int_{S_2} \mathcal{K}(h)\mathcal{K}(f)da_2,~~\forall f,h \in \{S_1\rightarrow \mathbb{R}\}$.
  \item Conformality holds, if and only if,   $~~\forall f,h \in \{S_1\rightarrow \mathbb{R}\},$
\\
 $$ \int_{S_1} \langle \nabla_{G_1}h,\nabla_{G_1}f\rangle_{G_1} da_1=\int_{S_2} \langle \nabla_{G_2}\mathcal{K}(h),\nabla_{G_2}\mathcal{K}(f)\rangle_{G_2}da_2,$$
 where $\nabla_{G_i}$ is the gradient with respect to the metric $G_i$ of $S_i$, $i=1,2$.
\end{itemize}

\subsection{Dirichlet energy and smooth maps}
\label{subsec:Dirichlet}
The smoothness of a map reflects its ability to map a smooth function to a smooth function.
One way to quantify the smoothness of such a map
 is to measure its Dirichlet energy.
\begin{defi}
\label{def:dirichlet}
The Dirichlet energy of the map $\mathcal{K}$ between two
 function spaces one on $S_1$ and the other on $S_2$, is defined by

$$
\begin{disarray}{ll}
 E_{\text{Dirichlet}} (\mathcal{K})&= \iint_{S_1,S_2}\|\nabla_{G_1(x)} k(x,y)\|^2da_1(x)da_2(y)\\
 &+ \iint_{S_1,S_2}\|\nabla_{G_2(y)} k(x,y)\|^2 da_1(x)da_2(y).
\end{disarray}
$$
\end{defi}
For any map $k(x,y)$, integration by parts yields
$$
\begin{disarray}{ll}
\int_{S_1}  \|\nabla_{G_1(x)} k(x,y)\|^2da_1(x)&=
 \int_{S_1}  \langle\Delta_{G_1(x)} k(x,y),k(x,y)\rangle da_1(x),
\end{disarray}
$$
where $\Delta_{G_i}$ represent the Laplace Beltrami operator of $S_i$.
Using these relations, we have
\begin{eqnarray*}
&\iint_{S_2,S_1} & \|\nabla_{G_2(y)} k(x,y)\|^2da_2(y)da_1(x) \cr
 &= &\iint_{S_2,S_1}  \langle\Delta_{G_2(y)} k(x,y),k(x,y)\rangle da_2(y)da_1(x)\cr
&\approx&\sum_i(\A_1)_{ii}\K_{i}^T\W_2\K_{i}=  \sum_i(\A_1)_{ii}\trace\left(\W_2\K_{i}\K_{i}^T\right)\cr
& = & \trace\left(\W_2\underbrace{\left(\sum_j\K_{j}\K_{j}^T(\A_1)_{jj}\right)}_{\K\A_1\K^T}\right)=\trace(\W_2\K\A_1\K^T),
\end{eqnarray*}
%
where $\W_i$ represents the cotangent weight matrix of the discretized Laplace Beltrami operator, $\Delta_{G_i} \approx \A_i^{-1} \W_i$,
 as introduced in \cite{cotan}.
We can similarly show that
$$
 \iint_{S_1,S_2}\|\nabla_{G_1(x)} k(x,y)\|^2 da_1(x)da_2(y)
 \approx \trace(\W_1\K^T\A_1\K).
$$
The discrete Dirichlet energy of a functional map can now be approximated by
\begin{equation}
\label{eq:dirichlet_energy}
E_{\text{Dirichlet}}
   (\K)=\trace(\W_2\K\A_1\K^T)+\trace(\W_1\K^T\A_2\K).
\end{equation}

\subsection{Mass Preservation}
\label{subsec:MP}
One of our requirements from the functional map is to be mass preserving.
 Formally, it has to satisfy
$$
\int_{S_2}\mathcal{K}(f_1)da_2=\int_{S_1}f_1da_1,
$$
 and
$$
\int_{S_1}\mathcal{K}^T(f_2)da_1=\int_{S_2}f_2da_2.
$$
Translating these conditions to matrix notations,
 the mass preservation property can be discretized into,
\begin{eqnarray}
\label{eq:mp}
\K\A_1\ones &=&\ones \cr
\K^T\A_2\ones &=&\ones.
\end{eqnarray}
where $\ones$ is vector whose components are all equal to one.
\subsection{Unitarity and local area preservation}
\label{subsec:Unitary}
An example of a functional unitarity is the Fourier transform.
Let $\mathcal{F}$ define the Fourier transform, then,
 we have that
$$
f=\mathcal{F}^*\left(\mathcal{F}(f)\right).
$$
Associating this property to
 the kernel in Equation (\ref{eq:continuous_func_map}),  allows
 us to write
$$
\begin{disarray}{ll}
 f_1(x)&=\int_{S_2}k(z,x)
    \underbrace{\left(\int_{S_1}k(z,\tilde{x})f_1(\tilde{x})da_1(\tilde{x})\right)}_{f_2(z)}da_2(z)\cr
 &=\iint_{S_1,S_2}k(z,x)k(z,\tilde{x})f_1(\tilde{x})da_1(\tilde{x})da_2(z),
\end{disarray}
$$
 and in a discrete setting,
\begin{equation}
\label{eq:unitarity}
\K^T\A_2\K\A_1=\I,
\end{equation}
 where $\I$ is the identity matrix.
This relation is equivalent to $\A_2\K\A_1\K^T=\I$, in which case,
 for any unitary map, we have,
\begin{eqnarray*}
 \K\A_1\K^T&=\A_2^{-1},\cr
 \K^T\A_2\K&=\A_1^{-1}.
\end{eqnarray*}
Plugging the above formulas into Equation (\ref{eq:dirichlet_energy}),
 it turns out that the Dirichlet energy of any unitary map is constant.
Moreover, if the map $\mathcal{K}$ is unitary, then, for all
 functions $f,h\in \{S_1\rightarrow \mathbb{R}\}$ we have,
 $$
\int_{S_2} \mathcal{K}(h)\mathcal{K}(f)da_2=\int_{S_1} h\mathcal{K}^*(\mathcal{K}(f))da_1=\int_{S_1} hfda_1.
 $$
This demonstrates the equivalence between a unitary map
 and a local area preserving one.

\subsection{Conformal map}
\label{subsec:conformal}
The conformality, also known as angular, or isotropy preserving,
 of a functional map $\mathcal{K}$ is equivalent to, 
 $$
 \begin{disarray}{l}
\int_{S_1} \langle \nabla_{G_1}h,\nabla_{G_1}f\rangle_{G_1} da_1=\int_{S_2} \langle \nabla_{G_2}\mathcal{K}(h),\nabla_{G_2}\mathcal{K}(f)\rangle_{G_2}da_2,
\\
\forall f,h \in \{S_1\rightarrow \mathbb{R}\}.
\end{disarray}
$$
Invoking Stockes theorem, the above equation can be written as
$$
\begin{disarray}{ll}
\int_{S_1} h\Delta_{G_1}fda_1&=\int_{S_2}\mathcal{K}(h)\Delta_{G_2}\mathcal{K}(f)da_2\\
&=\int_{S_1}h\mathcal{K}^*(\Delta_{G_2}\mathcal{K}(f))da_1,~~\forall f,h \in \{S_1\rightarrow \mathbb{R}\},
\end{disarray}
$$
that is equivalent to
$$
\Delta_{G_1}\cdot=\mathcal{K}^*(\Delta_{G_2}\mathcal{K}(\cdot)),
$$
or in discrete setting
\begin{equation}
\label{eq:conformality}
\A_1^{-1}\W_1=\K^T\A_2(\A_2^{-1}\W_2)\K\A_1=\K^T\W_2\K\A_1.
\end{equation}

\subsection{Eigenspace formulation}
\label{sec:eigen}
In \cite{fmap}, the authors define a discrete representation of the functional maps between
 shapes that involves the eigenspace of the discretized Laplace-Beltrami Operators (LBO) of $S_1$ and $S_2$.
Let $\PHI_i$ be the matrix that represents the eigenfunctions of the Laplace Beltrami operator of $S_i$ and $\LAMBDA_i$ its associated eigenvalues diagonal matrix, such that $\W_i\PHI_i=\A_i\PHI_i\LAMBDA_i$. The spectral representation of $\K$ with respect to $\PHI_1$ and $\PHI_2$ can be described by a matrix $\Alpha$ such that
$$
\K=\PHI_2\Alpha\PHI_1^T.
$$
In this setting, we readily have that,
$$
\K^T\A_2\K\A_1=\PHI_1\Alpha^T\PHI_2^T\A_2\PHI_2\Alpha\PHI_1^T\A_1.
$$
Now, since
$$
\PHI_2^T\A_2\PHI_2=\I,
$$
 then
$$
\K^T\A_2\K\A_1=\PHI_1\Alpha^T\Alpha\PHI_1^T\A_1.
$$
Now,  Condition (\ref{eq:unitarity}) can be written as
$$
\PHI_1\Alpha^T\Alpha\PHI_1^T\A_1=\I.
$$
Multiplying the left hand
 side by $\PHI_1^T\A_1$ and the right hand side by $\PHI_1$, given that
$$
\PHI_1^T\A_1\PHI_1=\I,
$$
 we conclude that Condition (\ref{eq:unitarity}) is simplified to
$$
\Alpha^T\Alpha=\I.
$$
Along the same line, the discrete Dirichlet energy (\ref{eq:dirichlet_energy})
 can be similarly simplified into\\
\begin{eqnarray*}
 E_{\text{Dirichlet}}(\K)&=&\trace(\W_2\K\A_1\K^T)+\trace(\W_1\K^T\A_2\K)\cr
&=&\trace(\W_2\PHI_2\Alpha\underbrace{\PHI_1^T\A_1\PHI_1}_{\I}\Alpha^T\PHI_2^T)\cr
&+&\trace(\W_1\PHI_1\Alpha^T\underbrace{\PHI_2^T\A_2\PHI_2}_{\I}\Alpha\PHI_1^T)\cr
&=&\trace(\W_2\PHI_2\Alpha\Alpha^T\PHI_2^T)+\trace(\W_1\PHI_1\Alpha^T\Alpha\PHI_1^T)\cr
&=&\trace(\Alpha\Alpha^T\underbrace{\PHI_2^T\W_2\PHI_2}_{\LAMBDA_2})
          +\trace(\Alpha^T\Alpha\underbrace{\PHI_1^T\W_1\PHI_1}_{\LAMBDA_1})\cr
&=&\trace(\Alpha\Alpha^T\LAMBDA_2)+\trace(\Alpha^T\Alpha\LAMBDA_1).
\end{eqnarray*}
The conformality equation (\ref{eq:conformality}) reads
$$
\A_1^{-1}\W_1=\K^T\W_2\K\A_1,
$$
and can be rewritten as
$$
\W_1=\A_1\PHI_1\Alpha^T\underbrace{\PHI_2^T\W_2\PHI_2}_{\LAMBDA_2}\Alpha\PHI_1^T\A_1,
$$
that is equivalent to
$$
\underbrace{\PHI_1^T\W_1\PHI_1}_{\LAMBDA_1}=\underbrace{\PHI_1^T\A_1\PHI_1}_{\I}\Alpha^T{\LAMBDA_2}\Alpha\underbrace{\PHI_1^T\A_1\PHI_1}_{\I},
$$
or
$$
\LAMBDA_1=\Alpha^T{\LAMBDA_2}\Alpha.
$$

Finally, the mass preservation, defined in Equation (\ref{eq:mp}),
  can be rewritten as
\begin{eqnarray*}
 \PHI_2\Alpha\PHI_1^T\A_1\ones &=&\ones \cr
 \PHI_1\Alpha^T\PHI_2^T\A_2\ones &=&\ones,
\end{eqnarray*}
 that is equivalent to
\begin{eqnarray}
\label{eq:mp_discrete}
\Alpha C_1&=&C_2\cr
\Alpha^T C_2&=&C_1,
\end{eqnarray}
where
$
C_i=\PHI_i^T\A_i\ones .
$

Putting all ingredients together, we consider spectral representation of smooth low area and angle distortion, mass preserving, linear maps,
 such that
\begin{enumerate}
\item $\K=\PHI_2\Alpha\PHI_1^T$,
\item $\|\Alpha^T\Alpha-\I\|$,  is as small as possible,
\item $\|\LAMBDA_1-\Alpha^T{\LAMBDA_2}\Alpha\|$ is as small as possible,
\item $\trace(\Alpha\Alpha^T\LAMBDA_2)+\trace(\Alpha^T\Alpha\LAMBDA_1)$ is as small as possible,
\item $\Alpha C_1=C_2$, and $\Alpha^T C_2=C_1$.
\end{enumerate}

 \section{Spectral interpolation}
 \label{sec:interp}
 Let us consider a triangulated surface $S$, with $n$ vertices $V_i$, and $\mathcal{J}$ a subset of $\{1,2,\ldots,n\}$
  such that $|\mathcal{J}|=m\leq n$. 

 Given a map $D:S\times S\rightarrow\mathbb{R}$ defined to every pair of points of $S$, and whose values are known
 at a given set of $m$ points $V_\mathcal{J}=\{V_j,j\in\mathcal{J}\}$, we can extend the value of $D$  by
 interpolating the value of $D$ over the other points of $S$, such that the map we get is as smooth as possible. Formally, we aim
 to  find a map $h$ defined on $S\times S$ whose values obtains at  $V_\mathcal{J}\times V_\mathcal{J}$ coincides with the values of $D$,
  and whose Dirichlet Energy introduced in Definition (\ref{def:dirichlet}) is minimal.  This problem of smooth interpolation could be written as
$$
\begin{disarray}{l}
\min_{h:S\rightarrow \mathbb{R}} E_\text{Dirichlet}(h)\\
\text{s.t.~~} h(V_i,V_j)=D(V_i,V_j)~~~~\forall (i,j) \in (\mathcal{J}\times\mathcal{J}).
\end{disarray}
$$
Using the spectral reformulation of this energy and defining by $\Alpha$ the spectral representation of $h$,  the problem can be rewritten as

\begin{equation}
\begin{disarray}{l}
\min_{\Alpha} \trace(\Alpha^T\LAMBDA\Alpha)+\trace(\Alpha\LAMBDA\Alpha^T)\\
\text{s.t.}~~(\PHI\Alpha\PHI^T)_{ij}=D(V_i,V_j),~~~~\forall(i,j)\in\mathcal{I},
\end{disarray}
\end{equation}
where $(\LAMBDA,\PHI)$ represent the diagonal matrices of eigenvalues and the matrix of eigenfunctions of the Laplace-Beltrami operator of $S$.
Expressing the constraint as a penalty function we end up with the following optimization problem
\begin{equation}
\label{eq:spectral_interp_prog_2}
\begin{disarray}{ll}
\min_{\Alpha\in \mathbb{R}^{m_e\times m_e}} &\trace(\Alpha^T\LAMBDA\Alpha)+\trace(\Alpha\LAMBDA\Alpha^T)\\
&+\mu\sum_{(i,j)\in\mathcal{I}}
     \|(\PHI\Alpha\PHI^T)_{ij}-D(V_i,V_j)\|_F^2,
     \end{disarray}
\end{equation}
 where $\|\cdot\|_F$ represents the Froebenius norm. 
 Problem (\ref{eq:spectral_interp_prog_2}) is a minimization problem of a quadratic function of $\Alpha$. 
Then, representing $\Alpha$ as an row-stack vector $\alpha$, the problem can be rewritten as a 
 quadratic programming problem. Next, let us recall the generalized multidimensional scaling procedure for shape matching, and then cast it into a spectral setting.

\section{GMDS}
\label{sec:GMDS}
Consider the shape correspondence problem that
 involves in searching for the best point to point assignment of
  two given shapes, $S_1$ and $S_2$.
The {\em Generalized Multi-Dimensional Scaling}   \cite{GMDS} is a procedure
 that computes the map that best preserves
 the inter-geodesic distances while embedding one surface into another.
Formally, if $\D_1$ and $\D_2$ represent the inter-geodesic distances matrix of $S_1$ and $S_2$, respectively,
 roughly speaking, the GMDS attempts to find the permutation matrix $\p$
   minimizing $\|\p\D_1-\D_2\p\|_2^2$.
It could be written as
\begin{eqnarray}
\label{eq:GMDS}
\min_{\p} &&\|\p\D_1-\D_2\p\|_2^2 \cr
\text{s.t.}&&\cr
&&\p\ \ones = \ones,\cr
&&\p^T \ones = \ones,\cr
&&\p_{ij}\in\{0,1\},~~~\forall (i,j).
\end{eqnarray}
It appears to be an NP hard problem that ignores the continuous nature of the shapes
 and their potentially smooth relation.
Several variations were proposed over the last years to reduce the intrinsic
 complexity of the problem \cite{GMDS,lipman,pokrass}.
Here, we start by following a similar initial path by
 relaxating the hard constraint $\p_{ij}\in\{0,1\},~~~\forall (i,j)$.
In addition, we  restrict 
 our solution to be unitary, mass and inter-geodesic distances preserving,
 with minimal conformal distortion, that produces a bijective linear map
   from $S_1$ to $S_2$, and defines a fuzzy correspondence between the surfaces.
Moreover, for the sake of consistency with the definition of a functional map,
 we replace $\p\X$ with $\p\A_1\X$ and $\X\p$ with $\X\A_2\p$.
Our new problem is defined by
\begin{eqnarray}
\label{eq:GMDS_relaxed}
\min_{\p} &&\|\p\A_1\D_1-\D_2\A_2\p\|_2^2 \cr
\text{s.t.}&&\cr
&&\p\A_1 \ones = \ones,\cr
&&\p^T\A_2 \ones = \ones,\cr
&&\p^T\A_2\p\A_1=\I, \cr
&&\|\W_1-\A_1\p^T\W_2\p\A_1\|<\epsilon.
\end{eqnarray}
 where $\|\cdot\|_2^2$ represents the discretization of the $L_2$ norm of a mapping 
 between $S_1$ and $S_2$.
In continuous setting,
$$
\|F\|_2^2=\iint_{S_1,S_2}F^2(x,y)da(x_1)da(x_2),
$$
 and in its discrete version,
$$
\|F\|_2^2\approx \tr{\F^T\A_2\F\A_1}.
$$
Then, our  $L_2$ measure defined in Equation (\ref{eq:GMDS_relaxed}) reads,
$$
\begin{disarray}{l}
\|\p\A_1\D_1-\D_2\A_2\p\|_2^2\\
=\tr{(\p\A_1\D_1-\D_2\A_2\p)^T\A_2(\p\A_1\D_1-\D_2\A_2\p)\A_1}\\
=-2\tr{\p^T\A_2\D_2\A_2\p\A_1\D_1\A_1}+C,
\end{disarray}
$$
 exploiting the relation $\p^T\A_2\p\A_1=\I$.

Then, Problem (\ref{eq:GMDS_relaxed}) can be reformulated as
\begin{eqnarray}
\label{eq:GMDS_relaxed_2}
 &\max_{\p}& \tr{\p^T\A_2\D_2\A_2\p\A_1\D_1\A_1} \cr
 &\text{s.t.}&\cr
 &&\p\A_1 \ones = \ones,\cr
 &&\p^T\A_2 \ones = \ones,\cr
 &&\p^T\A_2\p\A_1=\I,\cr
   &&\|\W_1-\A_1\p^T\W_2\p\A_1\|<\epsilon.
\end{eqnarray}
We are now ready to introduce smoothness to the game.

\section{Shape correspondence in spectral domain}
The correspondence $\p$ may be thought of as a functional map
 between $S_1$ and $S_2$, up to area normalization.
Thus, following the analysis in Section \ref{sec:functional_maps},
 we may write
\begin{equation}
\p = \PHI_2 \aalpha \PHI_1^T .
\end{equation}
The inverse operator is defined as
\begin{equation}
 \p^T = \PHI_1 \aalpha^T \PHI_2^T ,
\end{equation}
 so that  Property (\ref{eq:unitarity}), namely, $\p^T \A_2 \p \A_1 = \I$,
 holds for the correspondence map $\p$.
As shown in Section \ref{sec:functional_maps},
 this condition is equivalent to $\aalpha^T \aalpha = \I$.
In addition, for $\p$ to be  mass preserving (\ref{eq:mp}),
 we obtain similar constraints on the mapping $\aalpha$
\begin{eqnarray}
\aalpha \PHI_1^T \A_1 \ones 
    = \PHI_2^T \A_2 \ones,
    \\ \aalpha^T \PHI_2^T \A_2 \ones 
    = \PHI_1^T \A_1 \ones.
\end{eqnarray}

One of the important consequences of using the functional map representation
 of the correspondence is a reduction of the size of the problem.
We started by searching for a point-wise matching between the vertices of $S_1$ and those of $S_2$,
 with $\p \in [0, 1]^{|S_1| \times |S_2|}$.
Now, we consider the map $\aalpha$ relating between the bases $\PHI_1$ and $\PHI_2$, that is of size $M_1 \times M_2$,
 where $M_1\times M_2 \ll |S_1|\times |S_2|$.

Let us exploit the \emph{interpolated distance} representation introduced in \cite{spectral_MDS}, and briefly presented in 
Section \ref{sec:interp} according to which
\begin{eqnarray}
\label{eq:interpolated_dist}
 \tilde{\D}_i = \PHI_i \aalpha_i \PHI_i^T , & i = 1, 2.
\end{eqnarray}
Our target measure (\ref{eq:GMDS_relaxed_2}), now reads
$$
\begin{disarray}{l}
\max_\p\tr{\p^T\A_2\D_2\A_2\p\A_1\D_1\A_1} \cr
= \max_{\aalpha}\tr{\aalpha^T \aalpha_2 \aalpha \aalpha_1 \underbrace{\PHI_1^T \A_1 \PHI_1}_{=\I}} \cr
    = \max_{\aalpha}\tr{\aalpha^T \aalpha_2 \aalpha \aalpha_1}.
\end{disarray}
$$
We obtained a new optimization problem, where $\aalpha$ is our new argument.

\begin{eqnarray}
\label{eq:objective_alpha}
\max_{\aalpha} && \tr{\aalpha^T \aalpha_2 \aalpha \aalpha_1}\cr
\text{s.t.} &&\cr
&&\aalpha^T \aalpha \,= \I,\cr
&& \|\LAMBDA_1-\Alpha^T{\LAMBDA_2}\Alpha\| \,<\epsilon,\cr
&& \aalpha C_1 \,\,\,= C_2,\cr 
&& \aalpha^T C_2 = C_1 .
\end{eqnarray}
 where $ C_1 = \PHI_1^T \A_1 \ones$, and $C_2 = \PHI_2^T \A_2 \ones$.

Finally, we rewrite some of the constraints as penalty measures that yield
\begin{eqnarray}
\label{eq:objective_alpha}
\min_{\aalpha} && \|\aalpha \aalpha_1-\aalpha_2 \aalpha\|_2^2+\mu_1 \|\LAMBDA_1-\aalpha^T{\LAMBDA_2}\aalpha\|_2^2+\mu_2\|\aalpha^T \aalpha -\I\|_2^2 \cr
\text{s.t.} &&\cr
&& \aalpha C_1 \,\,\,= C_2,\cr 
&& \aalpha^T C_2 = C_1 .
\end{eqnarray}

\section{Experimental results}
Several experiments were performed in order to evaluate the accuracy and efficiency 
 of the proposed method.
We used 
 two publicly available datasets - TOSCA \cite{bronstein2008numerical} 
 and SCAPE \cite{anguelov2004correlated}. 
The TOSCA dataset contains $90$ densely sampled synthetic human and 
 animal surfaces, 
 divided into several classes with given point-to-point correspondences between 
 the shapes within each class.
The SCAPE dataset contains scans of real human bodies in different poses.

\begin{figure*}[htbp]
\centering
\includegraphics[width=1\columnwidth]{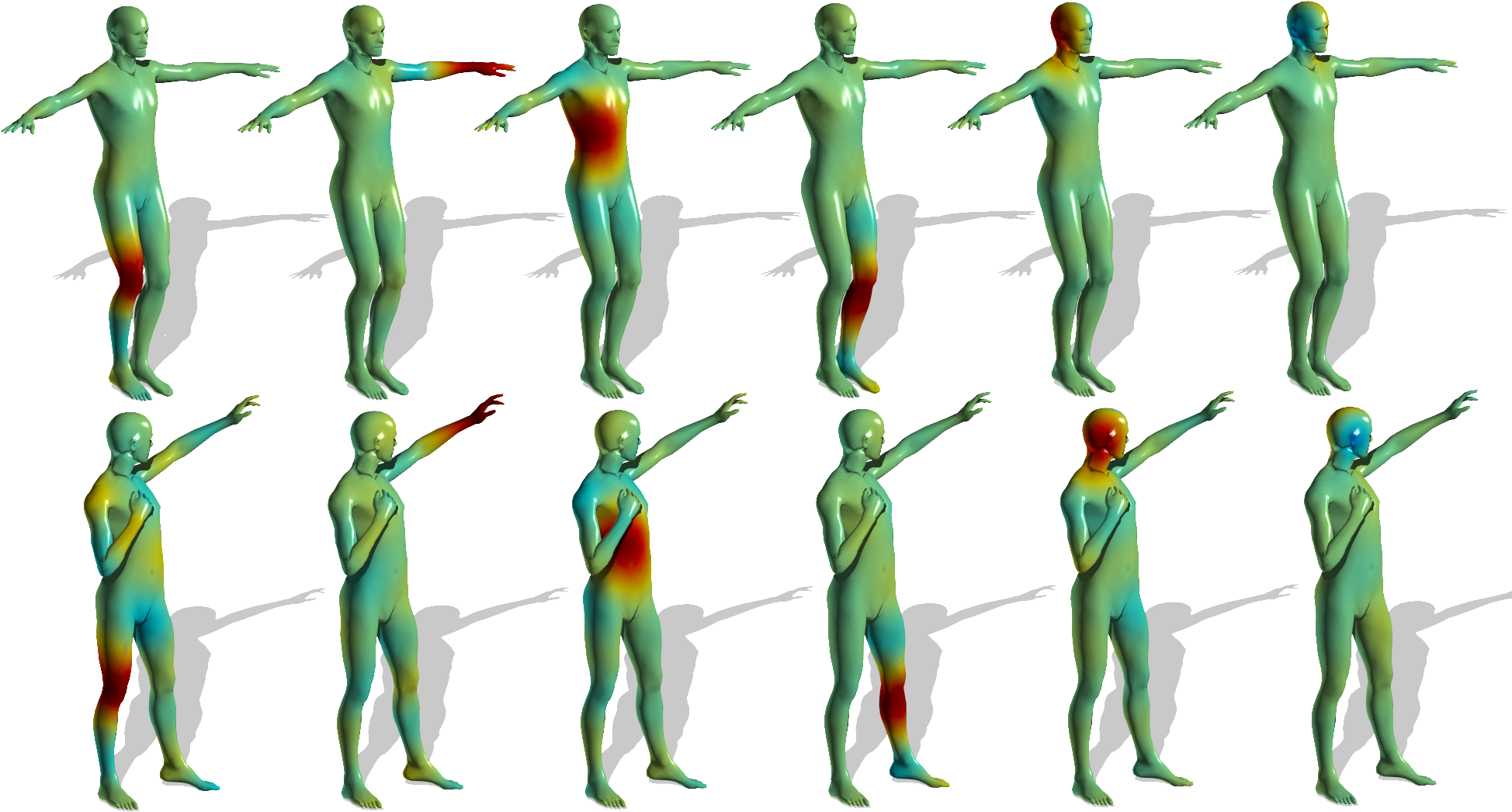}
\\[30pt]
\includegraphics[width=1\columnwidth]{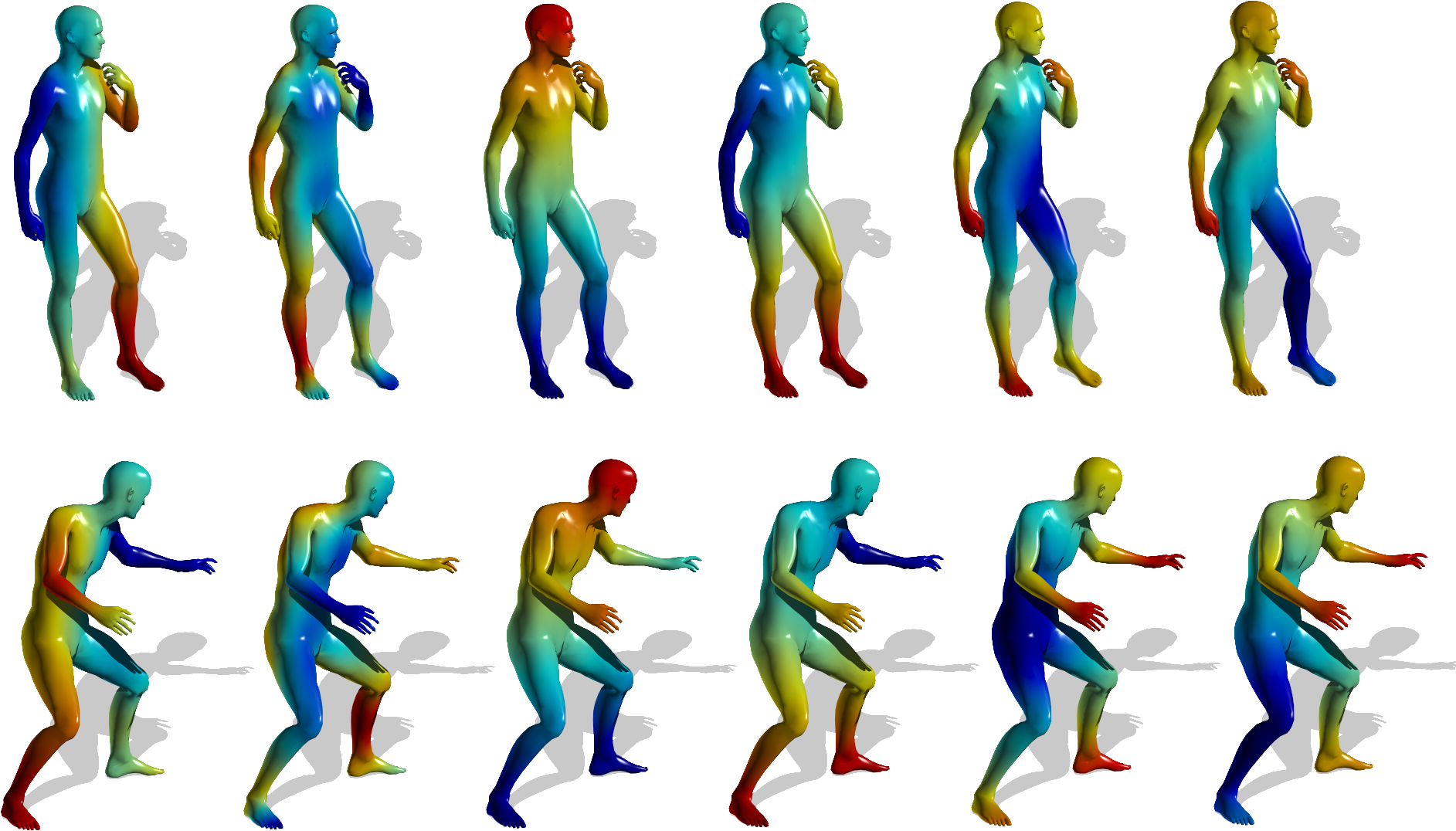}
\caption{Mapping  functions between two almost isometric shapes via S-GMDS.
}
\label{fig:simul_transfer_map1}
\end{figure*}

\begin{figure*}[htbp]
\centering
\includegraphics[width=1\columnwidth]{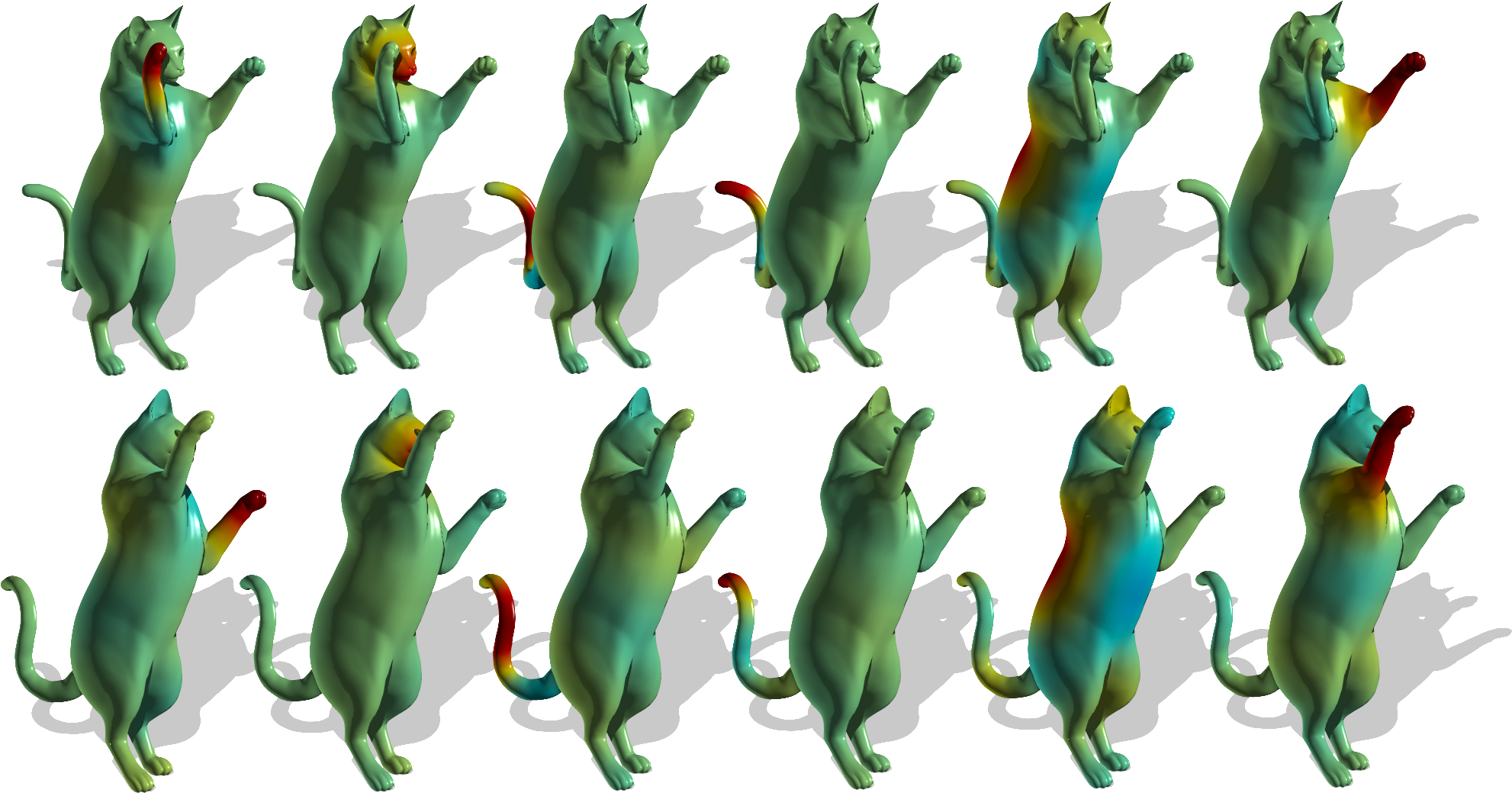}
\\[30pt]
\includegraphics[width=1\columnwidth]{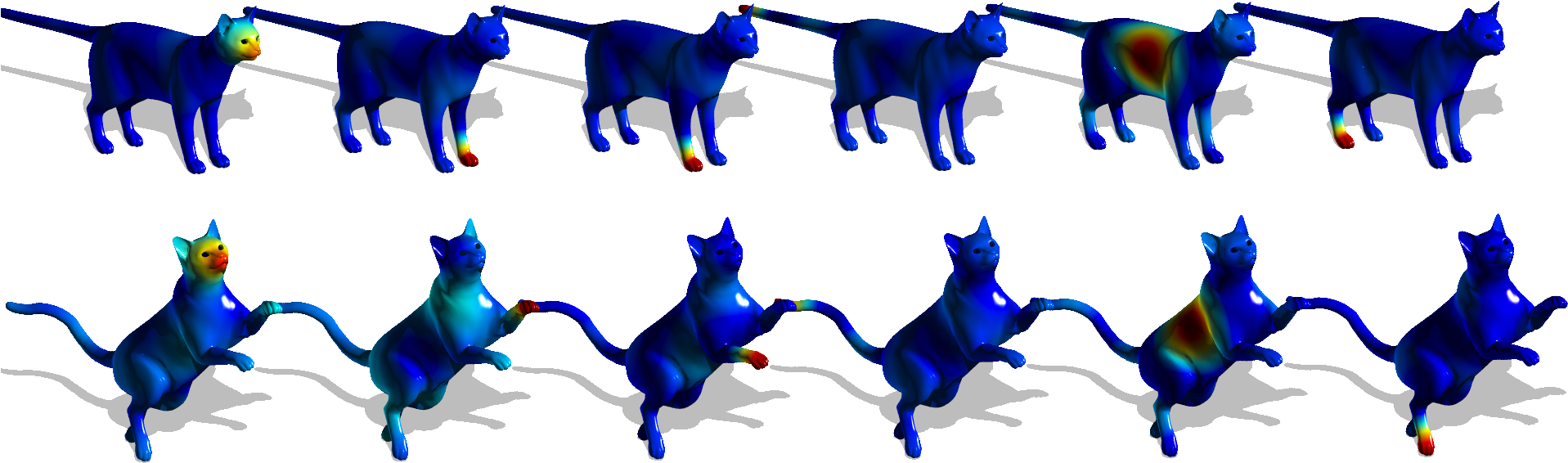}
\caption{Mapping functions between two almost isometric shapes via S-GMDS.}
\label{fig:simul_transfer_map2}
\end{figure*}

\begin{figure*}[htbp]
\centering
\includegraphics[width=0.28\columnwidth]{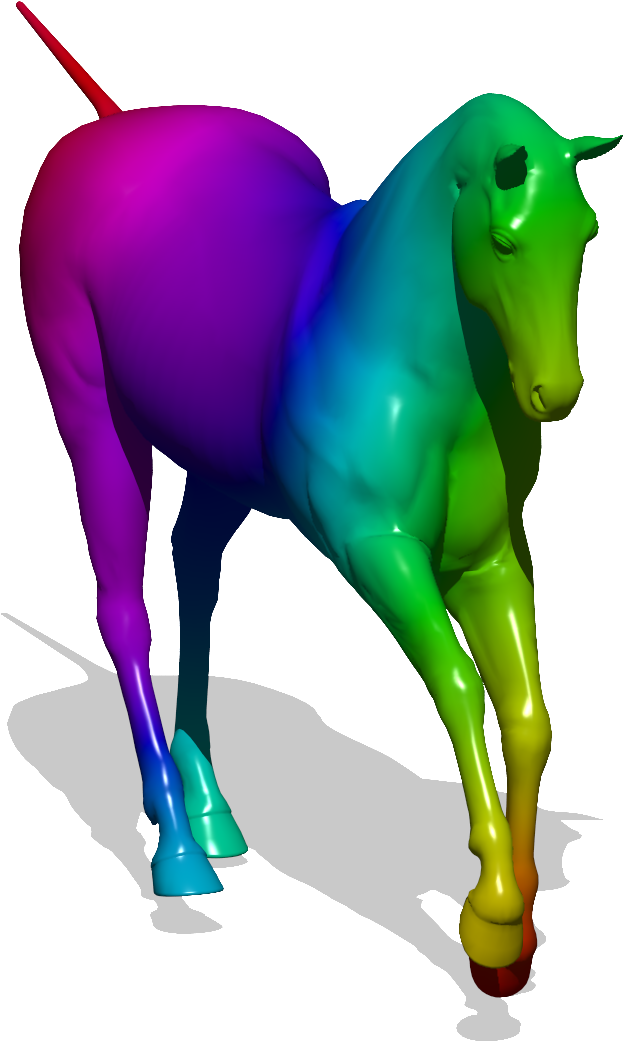}
\hfill
\includegraphics[width=0.28\columnwidth]{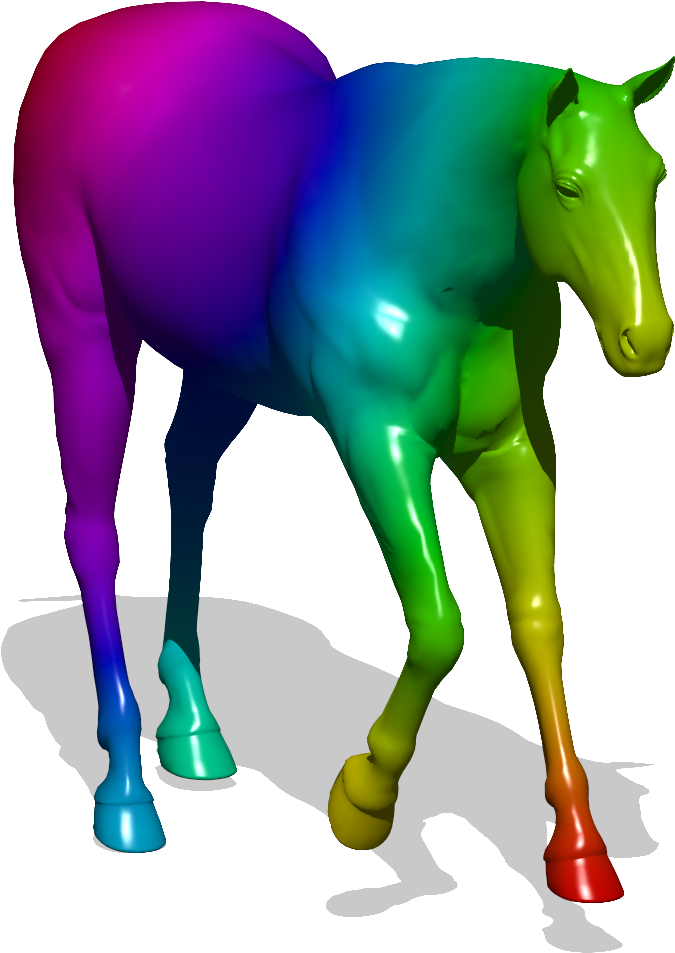}
\hfill
\includegraphics[width=0.28\columnwidth]{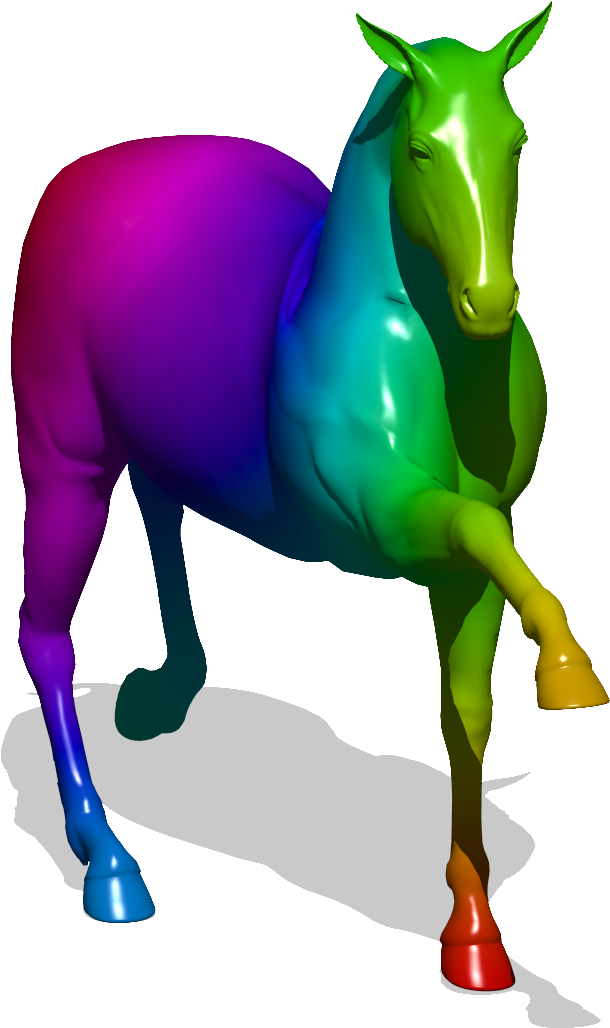}
\hfill
\includegraphics[width=0.28\columnwidth]{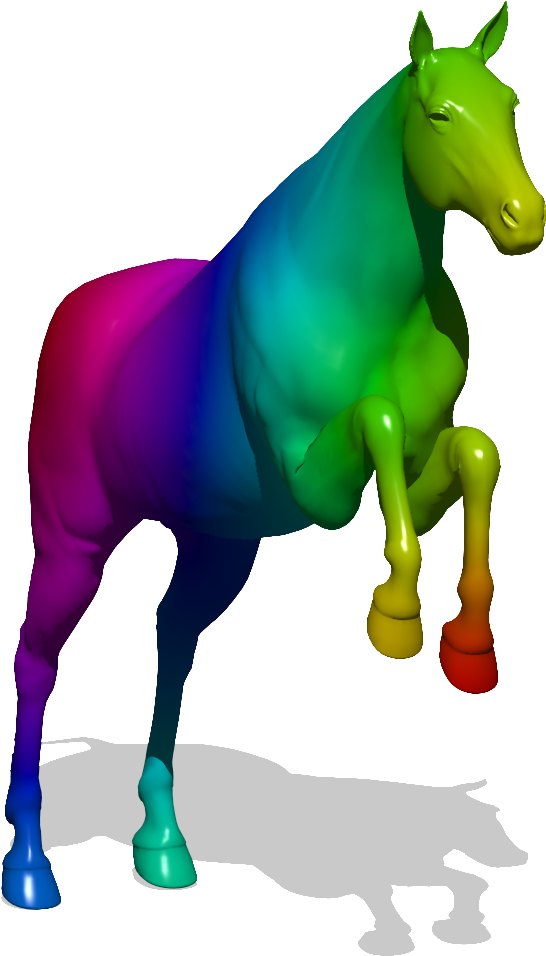}
\hfill
\includegraphics[width=0.28\columnwidth]{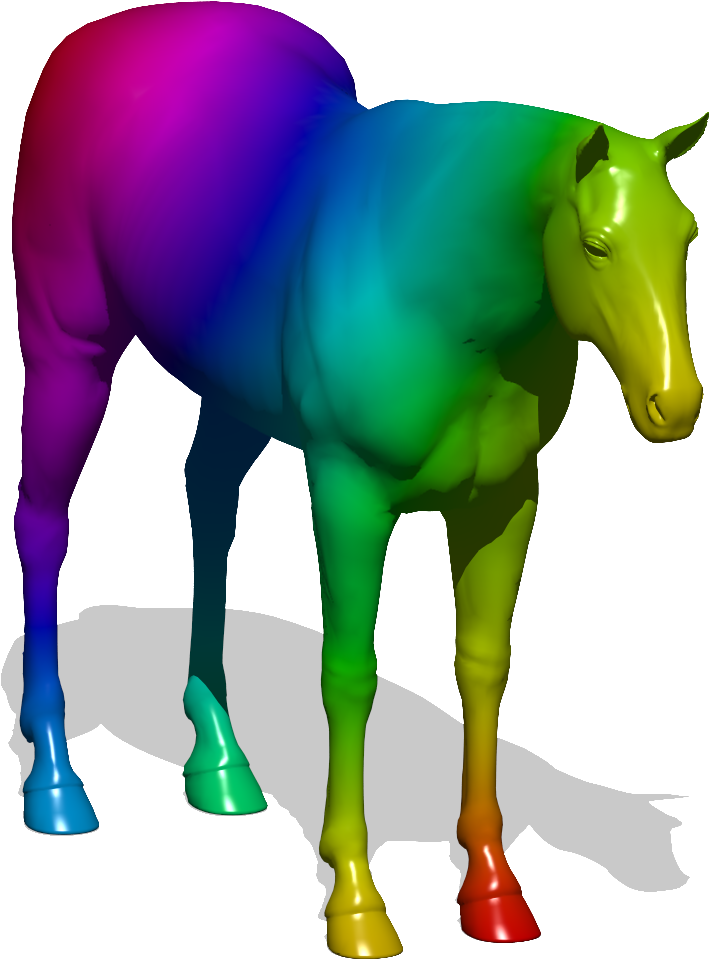}
\hfill
\includegraphics[width=0.28\columnwidth]{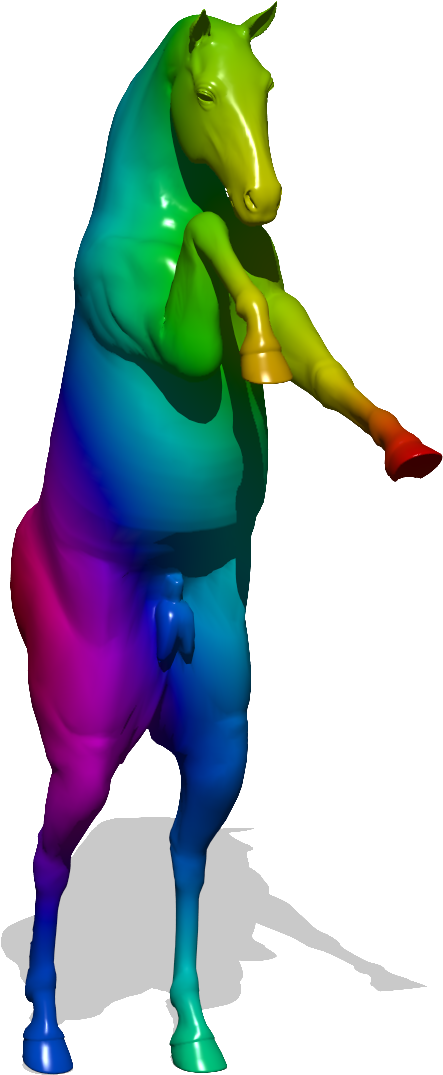}
\caption{Dense point-to-point correspondence between six almost isometric shapes 
 of a horse from the TOSCA dataset. 
}
\label{fig:simul_transfer_map_horse}
\end{figure*}

In our first experiment, we selected almost isometric surfaces within the same class
 from the TOSCA dataset, and computed correspondences between them 
 using the proposed Spectral-GMDS.
We visualize the quality of the mapping by transferring a couple of functions defined
 on one shape to the other, using the procedure from \cite{fmap}, as shown
 in Figures \ref{fig:simul_transfer_map1} and \ref{fig:simul_transfer_map2}.
In Figure \ref{fig:simul_transfer_map_horse} we visualize point-to-point correspondences 
 between several almost isometric poses of a horse, obtained using the S-GMDS.

\begin{figure}[htbp]
\centering
\includegraphics[width=1\columnwidth]{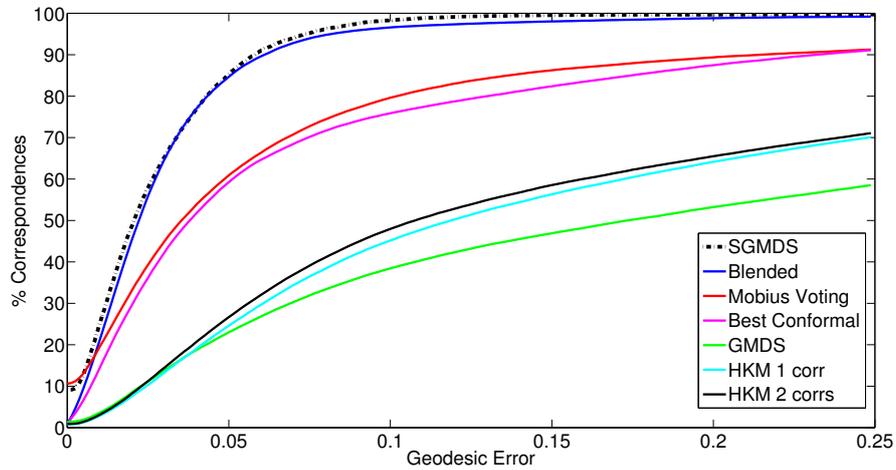}
\caption{Quantitative evaluation of the S-GMDS applied to shapes from the TOSCA dataset, 
 using the protocol from \cite{blended_map}.
}
\label{fig:exp_lipman_tosca}
\end{figure}

\begin{figure}[htbp]
\centering
\includegraphics[width=1\columnwidth]{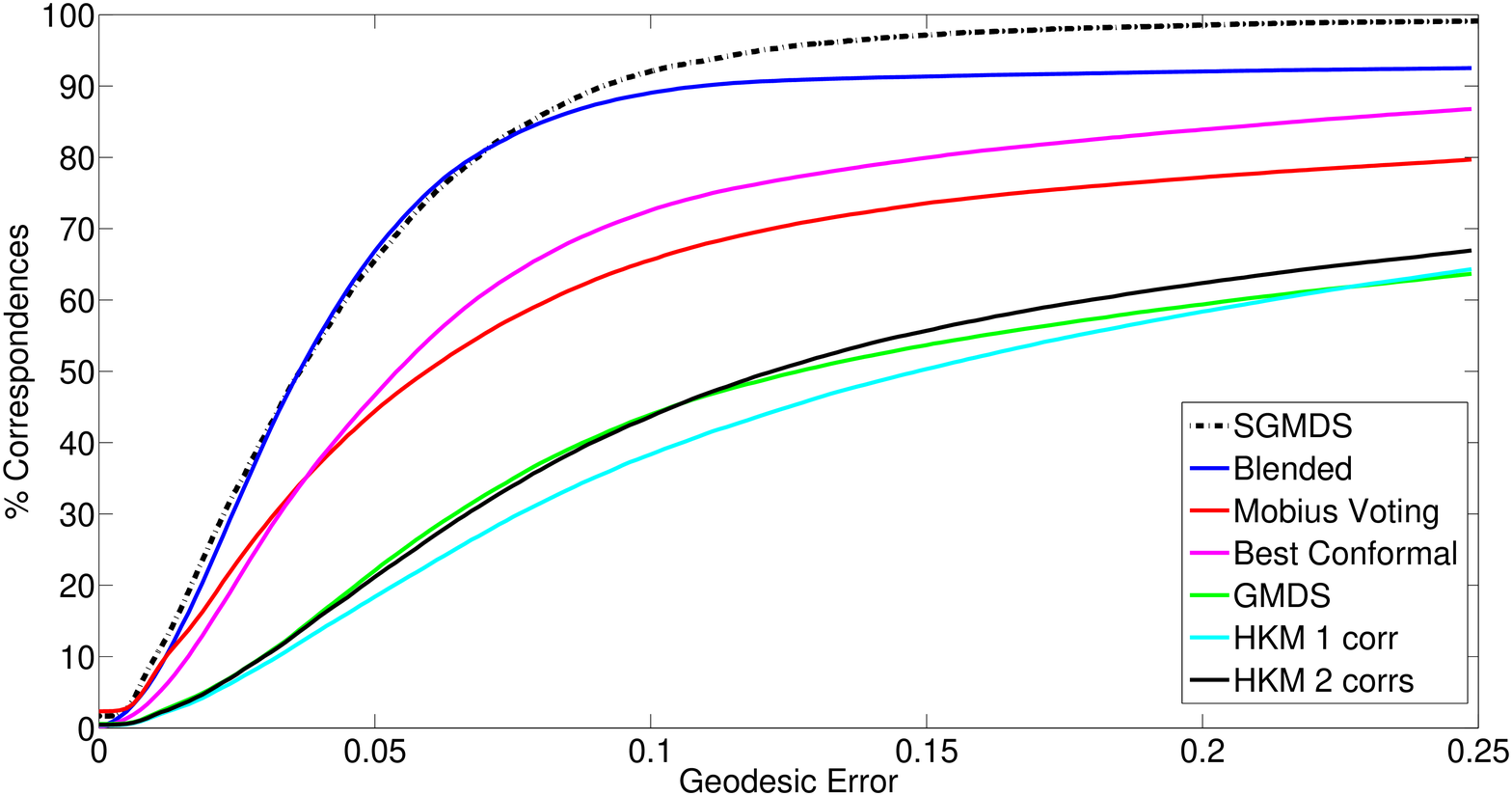}
\caption{Quantitative evaluation of the S-GMDS on shapes from the SCAPE dataset, 
  using the evaluation protocol from \cite{blended_map}.
   }
\label{fig:exp_lipman_scape}
\end{figure}

Figures \ref{fig:exp_lipman_tosca} and \ref{fig:exp_lipman_scape} compare the  
 accuracy of the proposed method to other methods using the evaluation 
  procedure proposed in \cite{blended_map}.
The evaluation protocol was applied to both TOSCA \cite{bronstein2008numerical}
 and SCAPE \cite{anguelov2004correlated} datasets.
For the other methods, we used the information provided in \cite{blended_map}.

In all of our experiments, we used pre-computed geodesic distances between a subset 
 of surface points, as defined in Equation (\ref{eq:interpolated_dist}).
The geodesic distances were calculated using the fast marching method \cite{Kimmel_Sethian:1998}, 
 between $5\%$ of surface points, sampled using the farthest point sampling 
  method \cite{hochbaum1985best,gonzalez1985clustering}.
To minimize the objective function in Equation (\ref{eq:objective_alpha}) we used the 
 PBM toolbox by M. Zibulevsky \cite{ben1997penalty}.
All the experiments were executed on a $2.7$ GHz Intel Core i7 machine with $16$GB RAM.
Average runtimes for pairs of shapes of various sizes from the TOSCA dataset are 
 shown in Table \ref{tbl:runtimes}. 
Figure \ref{fig:exp_noise},  demonstrates the robustness 
 of the proposed approach to typical types of noise.

In  the benchmark protocol proposed by Kim \etal \cite{blended_map}
 the  so-called ground-truth correspondence between shapes is assumed to be given.
Then, a script, provided by 
 the authors, computes the geodesic departure of each point, mapped by the evaluated
 method, from  what the authors refer to as true location.
The distortion curves describe the percentage of surface points falling within a
 relative geodesic distance from what is assumed to be their true locations.
For each shape, the geodesic distance is normalized with respect to the shape's squared root 
 of the area.
It is important to note that true location here is a subjective measure. 
In fact, measuring the geodesic distortion of the given
 correspondences demonstrates  a substantial discrepancy
 between corresponding pairs of points on most surface pairs 
 from the given datasets. 
The distortion curves would thereby have an intrinsic  ambiguity of about $5\%-25\%$. 
The state-of-the-art results reported in \cite{pokrass,fmap} thus reflect departure from 
 the isometric model, or over-fitting to the dataset or smooth interpolation between 
 corresponding features, 
 rather than the departure of the evaluated method from the isometry criterion.
The geodesic errors computed for the provided datasets could account for 
 subjective model fidelity rather than its axiomatic objective isometric accuracy. 
Based on Figure 6 in \cite{fmap}, the results by
 Kim \etal \cite{blended_map} could just as well be our best reference for state of the art.

\begin{figure}[htbp]
\mbox{
\hspace{-1cm}
\includegraphics[width=1\columnwidth]{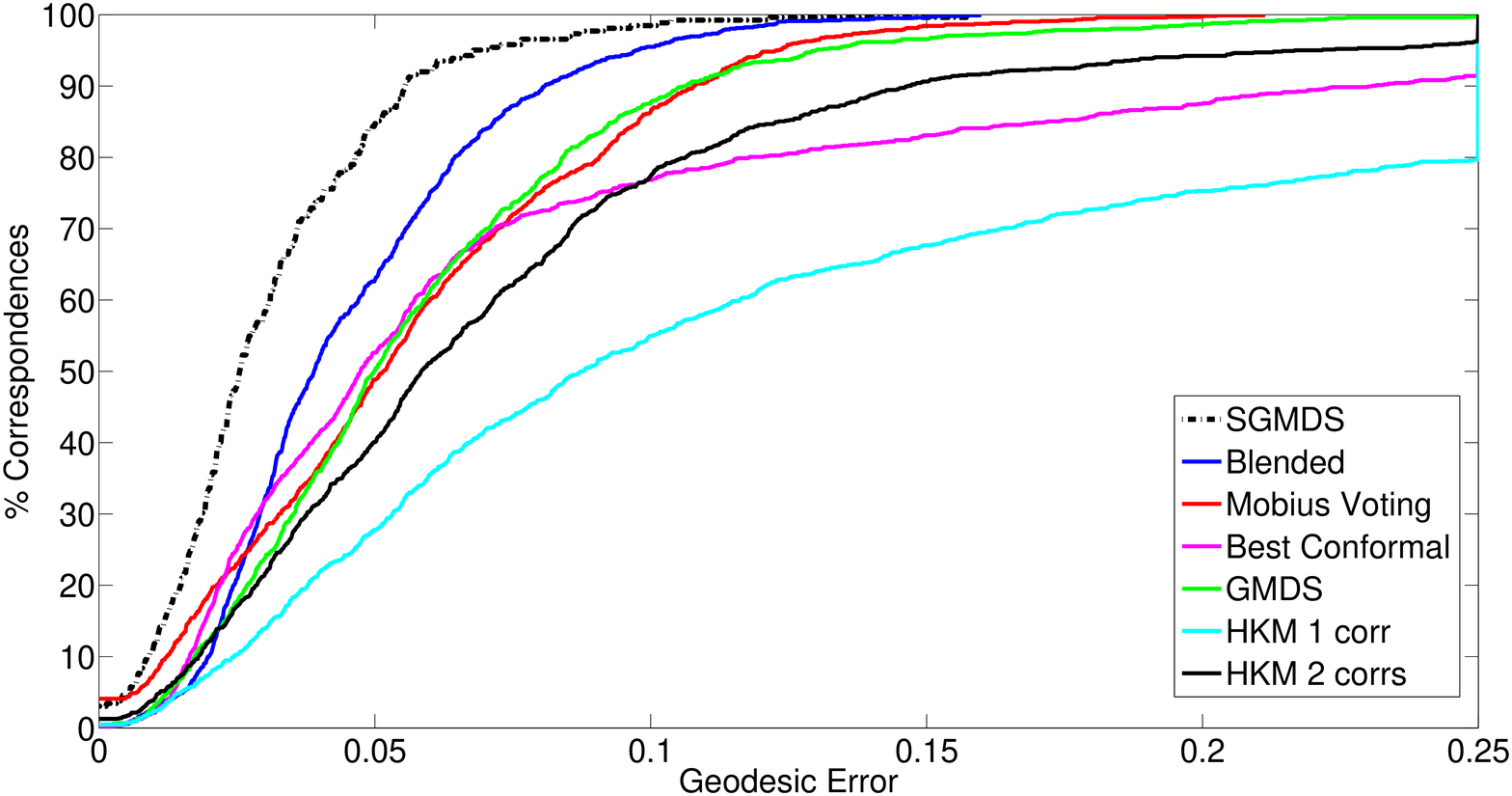}
\hspace{-0.7cm}
\includegraphics[width=0.25\columnwidth]{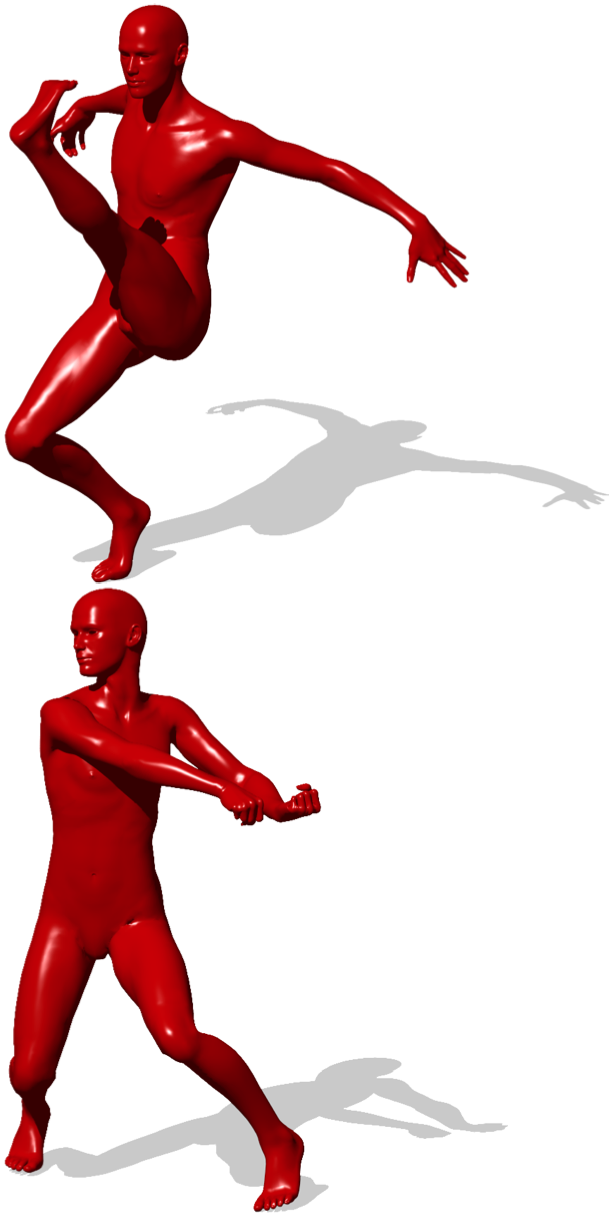} 
}
\caption{
 Performance evaluation of the S-GMDS compared to other methods
 applied to two ``David'' shapes from the TOSCA dataset which are relatively isometric.
 See shapes on the right.
 The comparison protocol is adopted from \cite{blended_map}.
 }
\label{fig:exp_lipman_david}
\end{figure}

\begin{figure*}[htbp]
\centering
\includegraphics[width=1\columnwidth]{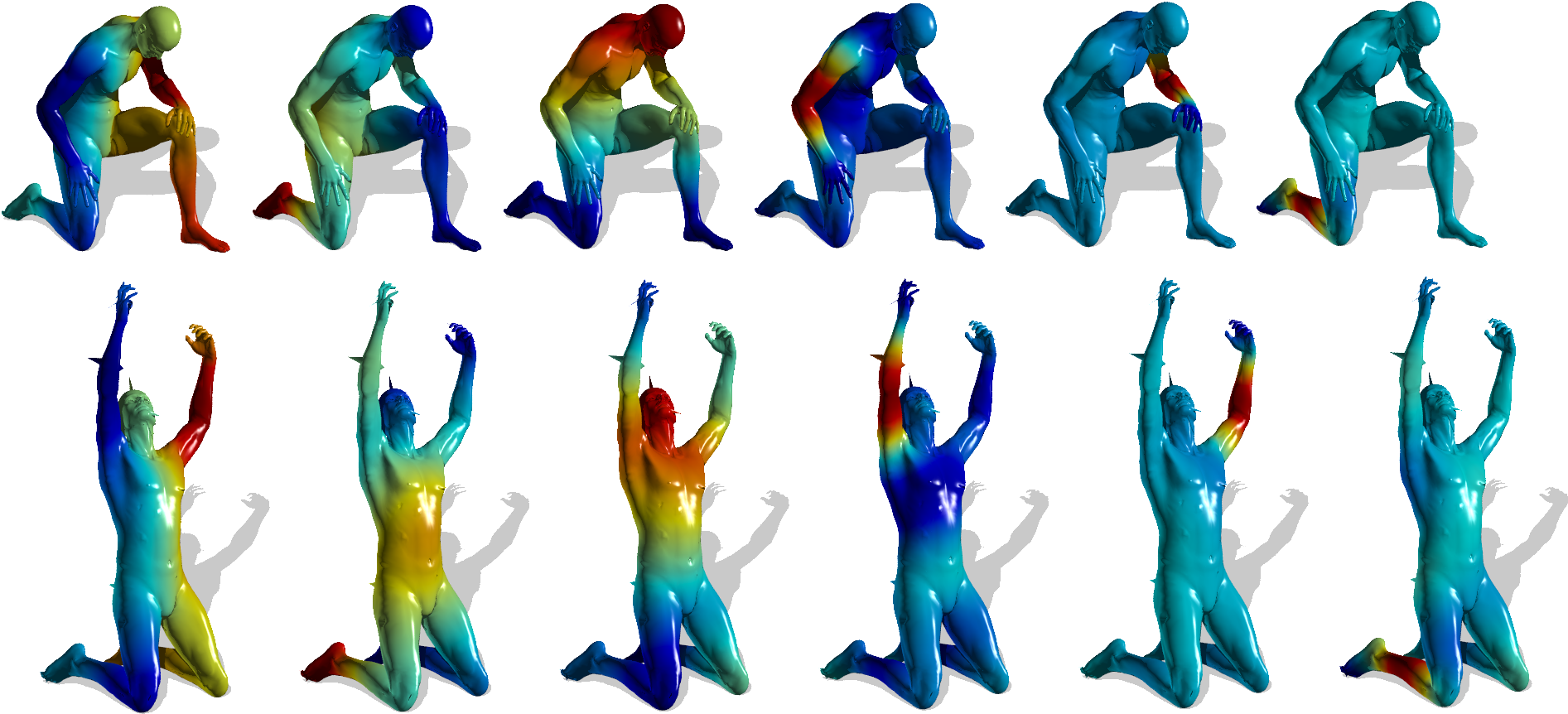}
\\[30pt]
\includegraphics[width=1\columnwidth]{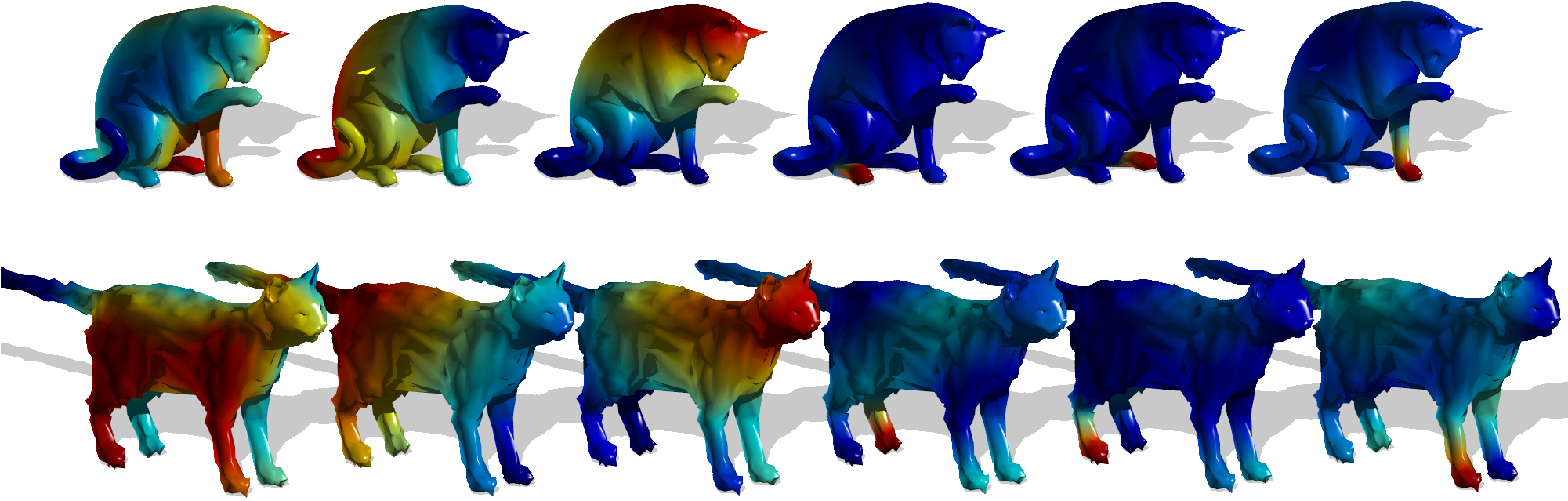}
\caption{Mapping  functions between two, almost isometric, noisy shapes via S-GMDS.}
\label{fig:exp_noise}
\end{figure*}

Still, even in this setting, the proposed method competes favorably 
 with state of the art results.
In a more favorable scenario,
 given two shapes for which the corresponding geodesic distortion is relatively small, 
 the S-GMDS provides superior results compared to existing methods, 
 as demonstrated in Figure \ref{fig:exp_lipman_david}. 
 


\begin{table}[t]
\tabcolsep7.5pt
\caption{Overall runtime (in seconds) of the proposed method 
 evaluated on shapes with various number of points (mesh vertices) 
 from the TOSCA dataset.\vspace{5pt}}{%
\begin{tabular}{@{}|c||c|c|c|c|c|c|c|c|@{}}
\hline
\textbf{\# Vertices} & 4344 & 19248 & 27894 & 45659 & 52565
\\ \hline
\textbf{\# Sampled vertices} & 217 & 962 & 1394 & 2282 & 2628
\\ \hline \hline
\textbf{LB + eigs} & 0.62 & 2.69 & 4.06 &6.43 & 7.47
\\ \hline
\textbf{Spectral GMDS} & 4.74 & 4.53 & 4.85 & 4.43 & 4.23
\\ \hline
\textbf{Total} & 5.36 & 7.22 & 8.92 & 10.86 & 11.71
\\
\hline
\end{tabular}}
\label{tbl:runtimes}
\end{table}
  
\section{Conclusions}
\label{sec:conclusion}
Spectral generalized multidimensional scaling method (S-GMDS) was proposed
 and proven to be an accurate model and efficient tool for matching non-rigid shapes. 
It accounts for almost isometric deformations of surfaces with respect to
 the regular metric.
Being able to account for distortions of large as well as small distances 
 when comparing two surfaces, with a natural regularization 
 of the matching, reduces the need for support of heuristics or initializations.
By incorporating the smoothness of the mapping, we treat the shape 
 matching problem holistically rather than as an interpolation between 
 multiple matched features such as points, regions, or localized functions.
  
Here, we used a regular metric in which geodesic distances on the surface 
 determine the isometric quantity we try to preserve and whose distortions we 
 use as a discrepancy measure. 
In our future research, we will try to axiomatically tackle the problem of 
 analyzing objects between which local scale can be 
 a substantial factor, yet, the conceptual meaning of such local structures with 
 different scale is preserved.
Though conformality could partially capture such distortions, we expect the 
 scale invariant geometry introduced in \cite{tech_report_aflalo_kimmel} and 
 plugged into the proposed framework to serve as the natural metric in such
 semi-local uniform scaling scenarios. 


\section{Acknowledgment}
The authors would like to thank Alon Shtern  for stimulating discussions and help with some of the computational tools.
This work has been supported by grant agreement no. 267414 of the European CommunityÕs FP7-ERC program.

\newpage

\bibliographystyle{plain}
\bibliography{SGMDS}


\end{document}